\documentclass[aps,prl,preprint]{revtex4-2}
\usepackage{graphicx}
\usepackage{epstopdf}
\usepackage{tabularx}
\usepackage{graphicx}
\usepackage{braket}
\usepackage{bm}
\usepackage{color, soul}
\usepackage{multirow}

\newcommand{\etal}{{\em et al}.\ }

\newcommand{\OP}{O$^{p}$}
\newcommand{\ONP}{O$^{np}$}

\begin{document}
\title{Emergent superconductivity in doped ferroelectric hafnia}
\author{Xu Duan}
\affiliation{Key Laboratory for Quantum Materials of Zhejiang Province, Department of Physics, School of Science, Westlake University, Hangzhou, Zhejiang 310024, China}
\author{Shi Liu}
\email{liushi@westlake.edu.cn}
\affiliation{Key Laboratory for Quantum Materials of Zhejiang Province, Department of Physics, School of Science, Westlake University, Hangzhou, Zhejiang 310024, China}
\affiliation{Institute of Natural Sciences, Westlake Institute for Advanced Study, Hangzhou, Zhejiang 310024, China}

\date{\today}

\newpage
\begin{abstract}{
Superconductivity and ferroelectricity, representing two distinct forms of ordered states, are typically not found together in the same system, making it even more difficult to create a connection between them. Here, supported by first-principles calculations, we propose that Anderson-Blount’s ferroelectric-like metal can be manifested in electron-doped ferroelectric $Pca2_1$ HfO$_2$. In this system, polar phonons and consequently ferroelectricity are not affected by the presence of itinerant electrons. We find that a nonpolar optical phonon, being strongly coupled to doped electrons, can acquire a pronounced electron-phonon coupling strength to activate conventional Bardeen-Cooper-Schrieffer superconductivity. The displacements of polar oxygen atoms in $Pca2_1$ HfO$_2$ create a link between ferroelectricity and superconductivity, enabling a tunable superconducting temperature ranging approximately from 10 to 30 Kelvin. Owing to hafnia's compatibility with silicon, we suggest HfO$_2$-based ferroelectric superconductors present an opportunity to construct high-performing hybrid integrated systems utilizing switchable quantum states.

}
\end{abstract}
\maketitle
\newpage

Superconductivity and ferroelectricity, despite their striking differences at the fundamental physics level, converge in a shared trait: the representation of ``ordered states"~\cite{Hwang12p103}. In superconductivity, the ordered state manifests through the orchestrated movement of coupled electron pairs, known as Cooper pairs, leading to the emergent phenomenon of zero electrical resistance. In contrast, the ordered state of a ferroelectric features the alignment of electric dipoles within the material, yielding a macroscopic polarization that can be manipulated by an external field. This symphony of internal order—the coordinated electron movement in superconductors and the collective response of electric dipoles in ferroelectrics— defines their distinctive and technologically useful properties~\cite{Dalven80p215,Holzman19p1800058,Scott07p954,Zhao18p054107}. However, reconciling these two distinct ordered states within a single material has been challenging become mobile electrons in metallic states tend to strongly screen the dipole-dipole interaction that is considered crucial for the emergence of ferroelectricity~\cite{Cohen92p136,Xiang14p094108}.  

A unique material system that concurrently exhibits both superconductivity and ferroelectricity is Sr$_{1-x}$Ca$_x$TiO$_{3-\delta}$ in which Ca dopants nudges the quantum paraelectric SrTiO$_3$ towards ferroelectricity, while oxygen vacancies provide electron-doping to induce superconductivity~\cite{Rischau17p643,Bednorz84p2289,Wang19p61}. The origin of superconductivity in doped SrTiO$_3$ and its interplay with ferroelectricity remains unclear~\cite{Edge15p247002,Kedem16p184507,vanderMarel19p013003,Kanasugi19p094504,Baratoff81p1335,Rischau17p643,Swartz18p1475,Tomioka19p738,Enderlein20p4852,Ruhman16p224515}. Multiple theories have been proposed to unravel this intricate phenomena with several of them emphasizing the importance of the soft transverse optical mode~\cite{Edge15p247002,Wolfle18p104505,vanderMarel19p013003}.
Based on their density functional theory (DFT) investigations of electron-doped ferroelectric BaTiO$_3$, Ma \etal~reported that the soft polar phonons are strongly modulated by the itinerant electrons~\cite{Ma21p2314}. This interaction could amplify the strength of electron-phonon coupling and induce phonon-mediated superconductivity. More recently, evidence of discrete switching of superconductivity at the ferroelectric transition was observed in bilayer T$_{\rm d}$-MoTe$_2$~\cite{Jindal23p48}. This discovery indicates that these two seemingly incompatible ordered states can indeed interact with each other, opening up opportunities for electrostatic control of quantum phases. 

The postulated superconductivity in electron-doped BaTiO$_3$ introduces a design principle for ferroelectric superconductors: utilizing a well-established ferroelectric through doping to exploit the strong coupling between soft polar phonons and doped itinerant carriers~\cite{Ma21p2314}. However, this strategy presents a fundamental compromise as an increased doping concentration often strongly suppresses the polar state~\cite{yang23p11721}. In this Letter, we demonstrate that it is feasible to realize the Anderson-Blount's weak electron coupling mechanism for ``ferroelectric-like metals"~\cite{Anderson65p217,Laurita19p3217,Puggioni14p3432}. This allows polar phonons to remain unaffected by charge-carrier doping, while nonpolar phonons displaying pronounced electron-phonon coupling can trigger conventional Bardeen-Cooper-Schrieffer (BCS) superconductivity.

\begin{figure}[ht]
\centering
\includegraphics[scale=0.5]{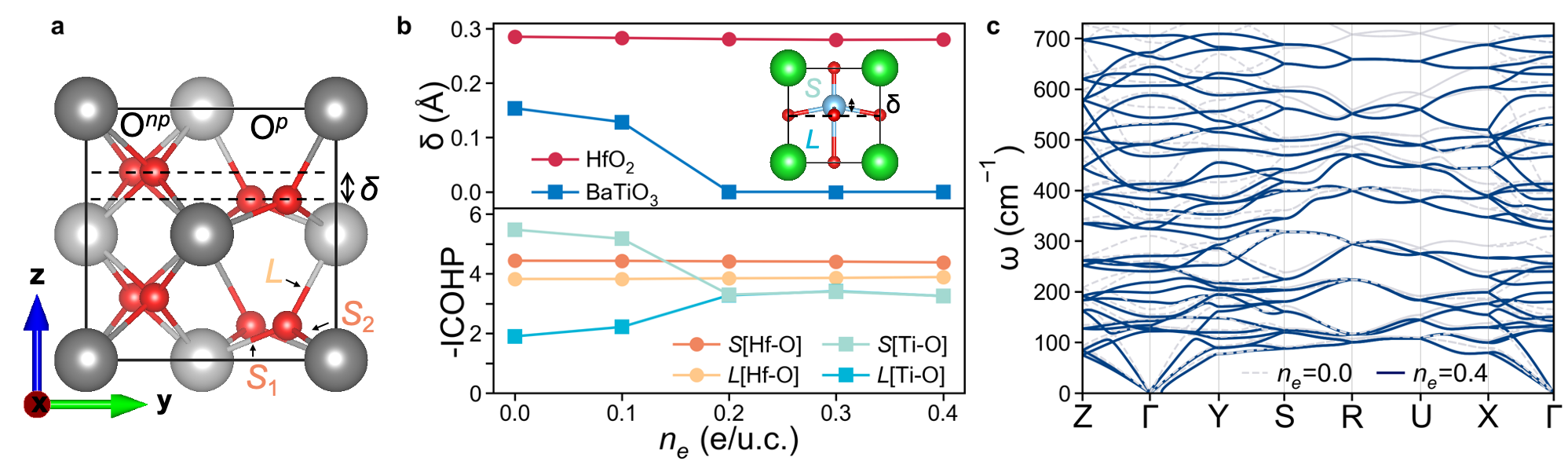}
 \caption{ \textbf{a} Crystal structure of ferroelectric $Pca2_1$ HfO$_2$. The displacement ($\delta$) of \OP~atoms along the $z$-axis breaks the inversion symmetry. The \OP~atoms are coordinated to three Hf atoms, forming three bonds denoted as $L$, $S_1$, and $S_2$, respectively.  
 (b) Polar atomic displacement $\delta$ and negative integrated Crystal Orbital Hamiltonian Population ($-$ICOHP) as a function of electron-doping concentration ($n_e$) for $Pca2_1$ HfO$2$ and tetragonal BaTiO$_3$. The inset illustrates the polar displacement of Ti in the unit cell of tetragonal BaTiO$_3$. The bottom panel shows the averaged $-$ICOHP value of $S_1$[Hf-O] and $S_2$[Hf-O] as $S$[Hf-O]. \textbf{c} Phonon spectra for undoped ($n_e=0.0$) and electron-doped ($n_e=0.4$) $Pca2_1$ HfO$_2$.}
  \label{fig1}
 \end{figure}

The model system we identified is electron-doped ferroelectric hafnia (HfO$_2$). 
Recognized by the industry for their silicon compatibility, HfO$_2$-based ferroelectrics emerge as an outstanding candidate for integrating ferroelectric functionalities into silicon-based circuits~\cite{Boscke11p102903,Kim21peabe1341}. In contrast to conventional perovskite ferroelectrics, HfO$_2$-based thin films maintain robust switchable polarization even at nanometer thickness~\cite{Chernikova16p7232,Lee20p1343,Cheema20p478}. 
Despite the hotly debated origin of ferroelectricity in hafnia thin films, the polar orthorhombic $Pca2_1$ phase is often identified as the ferroelectric phase~\cite{Park15p1811,Huan14p064111,Sang15p162905,Materlik15p134109}. 
The unit cell of $Pca2_1$ HfO$_2$ (Fig.~\ref{fig1}\textbf{a}) possesses a spacing layer consisted of fourfold-coordinated nonpolar oxygen ions (\ONP) that separates threefold-coordinated polar oxygen ions (\OP). Previous DFT studies have shown that the $Pca2_1$ phase exhibits an extreme charge-carrier-inert ferroelectricity, where local polar displacements of \OP~atoms are insensitive to charge-carrier doping~\cite{Ma23p096801}. This unique feature makes doped HfO$_2$ an ideal platform to investigate doping-induced superconductivity with minimal disruption to the polar state.

Here, supported by DFT calculations, we find that electron-doped ferroelectric HfO$_2$ can transition into a BCS superconductor with moderate superconducting transition temperatures ($T_c$) in the range of tens of Kelvins. Unlike  BaTiO$_3$ that experiences an increase in electron-phonon coupling strength due to doping-induced polar mode softening, the enhanced $T_c$ in electron-doped HfO$_2$ primarily stems from the softening of a nonpolar optical mode. By leveraging the interaction between this nonpolar phonon and the polar displacements of \OP~atoms, we demonstrate a switchable $T_c$ ranging approximately from 10 to 30 K. The silicon-compatible ferroelectric superconductor proposed here presents opportunities to integrate superconducting devices with existing silicon-based electronics. This could pave the way for scalable hybrid systems with novel functionalities powered by switchable quantum phases.

DFT calculations are carried out using QUANTUM ESPRESSO~\cite{Giannozzi09p395502,Giannozzi17p465901} with local density approximation (LDA)~\cite{Perdew81p5048} and pseudopotentials taken from the Garrity-Bennett-Rabe-Vanderbilt (GBRV) pseudopotential library~\cite{Garrity14p446}.
The structural parameters of the unit cell of $Pca2_1$ HfO$_2$ are optimized using a plane-wave cutoff of 90 Ry, a charge-density cutoff of 720 Ry, 9$\times$9$\times$9 $\Gamma$-centered \textbf{k}-point mesh for Brillouin zone (BZ) sampling, 
an energy convergence threshold of 10$^{-7}$ Ry, a
force convergence threshold of 10$^{-6}$ Ry/Bohr, and a stress threshold of 0.5 kBar.  The LDA lattice constants are comparable with experimental values.
We note that rather large plane-wave and charge density cutoffs are employed to achieve convergence in the electronic structure, particularly for the doped systems.  
Following the protocol in previous studies, the impacts of electrostatic doping are investigated with the background-charge approach with fixed volume in which the number of electrons is adjusted to set the doping concentration while an additional homogeneous background charge is introduced to ensure charge neutrality~\cite{Michel21p8640,Bruneval15p024107,Daniel23p010301}. For a specific electron-doping concentration ($n_e$) in the unit of electron per unit cell ($e$/u.c.), the atomic positions are fully relaxed with lattice constants fixed to undoped ground-state values, followed by a single-point energy calculation using a tighter threshold for self-consistency (10$^{-12}$ Ry).
The phonon spectrum is calculated with the density functional perturbation theory using a 3$\times$3$\times$3 \textbf{q}-point mesh.  Maximally localized Wannier functions (MLWFs) along with the Migdal-Eliashberg theory implemented in Wannier90~\cite{Marzari12p1419,Pizzi20p165902} and EPW~\cite{Ponc16p116,Giustino07p165108} are used to quantify the electron-phonon coupling strength. To begin, the electron-phonon matrix elements are computed using a 6$\times$6$\times$6  \textbf{k}-point mesh in the electron BZ and a 3$\times$3$\times$3 \textbf{q}-point mesh in the phonon BZ. These elements are subsequently interpolated onto finer grids (18$\times$18$\times$18 for \textbf{k}/\textbf{q} points) using MLWFs. The chemical bonding situations in electron-doped HfO$_2$ are analyzed with the integrated Crystal Orbital Hamilton Population
(ICOHP) method implemented in LOBSTER~\cite{Dronskowski93p8617,Maintz16p1030}. All structural files and some representative input
files for DFT calculations are uploaded to a public repository.

The spontaneous polarization in $Pca2_1$ HfO$_2$ results from the atomic displacements ($\delta$) of \OP~atoms relative to the center of neighboring Hf planes along the $z$-axis (Fig.~\ref{fig1}\textbf{a}). Figure~\ref{fig1}\textbf{b} illustrates the value of $\delta$ as a function of $n_e$ together with the results for BaTiO$_3$ included for comparison. Consistent with previous DFT studies~\cite{Michel21p8640,Ma21p2314}, electron-doping drastically suppresses the polar displacements of Ti atoms in BaTiO$_3$ but induces a negligible impact on the magnitude of $\delta$ in $Pca2_1$ HfO$_2$. We further compute ICOHPs for the three Hf-\OP~bonds and find that they preserve nearly constant bonding strengths, showing no significant sensitivity to increased electron-doping (Fig.~\ref{fig1}\textbf{b} bottom panel).  
This behaviour contrasts with that observed in BaTiO$_3$: electron-doping weakens the shorter Ti-O bond while strengthening the longer Ti-O bond along the polar axis, consequently diminishing the extent of inversion symmetry breaking. Additionally, a high doping concentration of $n_e=0.4$ does not compromise the dynamical stability of $Pca2_1$ HfO$_2$, as confirmed by the corresponding phonon spectrum that exhibits no imaginary frequencies along high-symmetry lines of the BZ (Fig.~\ref{fig1}\textbf{c}).

\begin{figure}[ht]
\centering
\includegraphics[scale=0.5]{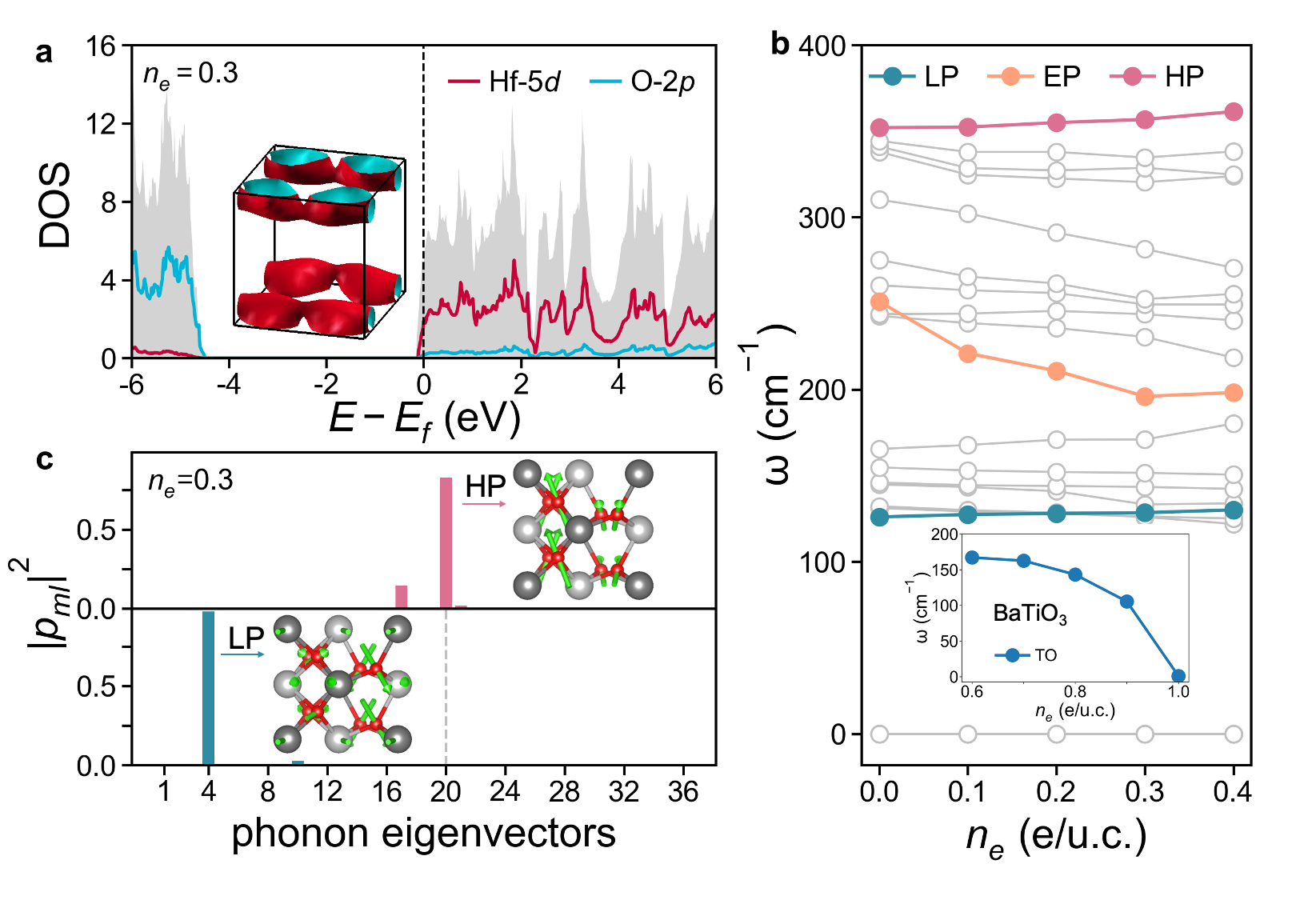}
 \caption{\textbf{a} Projected density of states (DOS) of electron-doped $Pca2_1$ HfO$_2$ at $n_e=0.3$. The inset shows the Fermi surface.
 \textbf{b} $\Gamma$-point phonon frequencies as a function of $n_e$. Blue and red filled circles highlight two polar optical phonons, LP and HP, respectively. The doping-induced softening of a zone-center polar transverse optical (TO) mode in tetragonal BaTiO$_3$ is depicted in the inset.  
 \textbf{c} Projection coefficients  ${|{p_{ml}}|^2}$ for HP (top) and LP (bottom) at $n_e=0.3$ using phonon eigenvectors of undoped HfO$_2$ as the basis set. Insets show the eigenvectors of undoped HP and LP. }
  \label{fig2}
 \end{figure}

The origin of the extreme charge-carrier-inert ferroelectricity in $Pca2_1$ HfO$_2$ is elusive. In its pristine state, ferroelectric HfO$_2$ is an insulator with a LDA bandgap of approximately 4.32 eV. Upon adding electrons, the Fermi level rises through the conduction band, transforming the system into a doped metal. The projected density of state (DOS) analysis indicates that the added electrons will predominantly occupy Hf-$5d$ orbitals (Fig.~\ref{fig2}\textbf{a}). Comparatively in electron-doped BaTiO$_3$, the added electrons will mostly be accommodated in Ti-$3d$ orbitals, causing Ti cations to deviate from the formally $d^0$ electron configuration and reducing the second-order Jahn-Teller (SOJT) activity responsible for spontaneous symmetry breaking~\cite{Michel21p8640}. In the case of $Pca2_1$ HfO$_2$, the inversion symmetry breaking has been ascribed to a trilinear coupling of three modes, similar to that in hybrid improper ferroelectrics~\cite{Delodovici21p064405}. Structurally, the polarization of HfO$_2$ originates from the displacements of anions rather than cations as in BaTiO$_3$. Therefore, it is plausible to hypothesize that the ferroelectricity in $Pca2_1$ doesn't  involve SOJT of Hf cations and is less affected by electron-doping. This is also confirmed by the weak $n_e$-dependence of the barrier height that separates two opposite polar states of $Pca2_1$ HfO$_2$ (see Supplemental Material).

To gain a microscopic understanding of the doping effect, we calculate the phonon frequencies at the $\Gamma$ point as a function of $n_e$. The results for modes with frequencies below 380 cm$^{-1}$ are illustrated in Fig.~\ref{fig2}\textbf{b}. As electron-doping can affect mode frequencies and potentially induce mode mixing, establishing a correlation between modes at different doping concentrations can be difficult. We use the mode-projection method~\cite{Yimer20p174105,Raeliarijaona15p094303} to track the evolution of a phonon mode with varying $n_e$. At the BZ center, since the phonon eigenvectors at one doping concentration $n_{e1}$ serve as a complete basis set, they can be applied to expand the phonon eigenvectors at anther doping concentration $n_{e2}$ as
\begin{equation}
	\ket{ \epsilon_l^{i\alpha}(n_{e2})} = \sum_m p_{ml} \ket{\epsilon_m^{i\alpha}(n_{e1})}.
\end{equation}
Here, $\ket{\epsilon_l^{i\alpha}(n_{e1})}$ and  $\ket{\epsilon_m^{i\alpha}(n_{e2})}$ represent the phonon eigenvectors at doping concentrations $n_{e1}$ and $n_{e2}$, respectively; $m$ and $l$ are mode indices, $i$ is the atom index, and $\alpha$ specifies the direction ($x$, $y$, $z$). The mode-projection coefficient, $p_{ml}$, characterizing the correlation between mode $m$ and mode $l$, is given as 
\begin{equation}
	p_{ml}=\sum_{i\alpha} \langle       \epsilon_m^{i\alpha}(n_{e1})\left| \epsilon_l^{i\alpha}(n_{e2})\right. \rangle.
\end{equation}
Based on the values of $p_{ml}$, we relate the phonon modes between two neighboring $n_e$, as represented by the solid lines in Fig.~\ref{fig2}\textbf{b}.

We first discuss the electron-doping effect on two flat polar phonon modes in $Pca2_1$ HfO$_2$ identified in previous studies~\cite{Lee20p1343}. The low-frequency polar mode (denoted as LP) involves atomic displacements of all the modes that undergo condensation during the cubic-to-orthorhombic transition, while the high-frequency polar mode (referred to as HP) involves  polar and antipolar displacements of oxygen atoms. Figure~\ref{fig2}\textbf{c} displays the projection coefficients 
$|p_{ml}|^2$ for LP and HP at $n_e=0.3$, using the phonon eigenvectors of undoped HfO$_2$ as the basis set.  
In the undoped HfO$_2$, LP corresponds to the lowest-frequency optical model ($m=4$). At $n_e=0.3$, LP is characterized by $l=5$ and $|p_{m=4,l}|^2=0.9$.
This means despite the reordering of modes based on frequencies after doping, the phonon eigenvector of LP largely retains its original undoped form. In comparison, 
the doped HP is identified by $l=20$ and is derived from two undoped modes, 15\% from a mode with $m=17$ and 85\% from the original HP ($m=20$). Overall, electron-doping does not cause substantial mode mixing for the two polar modes in ferroelectric HfO$_2$. This markedly differs from perovskite ferroelectrics where the mode mixing caused by metallicity is a fairly universal phenomena and occur in many systems such as BaTiO$_3$, PbTiO$_3$~\cite{Yimer20p174105} and LaSrMnO$_3$/LaNiO$_3$~\cite{Ghosh17p177603}. Another feature of HfO$_2$ distinct from BaTiO$_3$ is the doping-induced hardening of polar modes: the frequency of LP (HP) depends on $n_e$ linearly with a positive slop ($\Delta \omega / \Delta n_e$) of 10.3 cm$^{-1}$ (23.4 cm$^{-1}$), as shown in Fig.~\ref{fig2}\textbf{b}. These results suggest that electron-doped $Pca2_1$ HfO$_2$ qualifies as an Anderson-Blount ferroelectric metal in which polar modes, primarily involving vibrations of O atoms,
are weakly coupled to itinerant electrons located on Hf atoms. A similar weak coupling feature likely presents in polar metals like LiOsO$_3$, where the free carriers are situated on Os and O atoms while the Li atoms contribute to most polar distortions~\cite{Laurita19p3217}.

\begin{figure}[ht]
\centering
\includegraphics[scale=0.5]{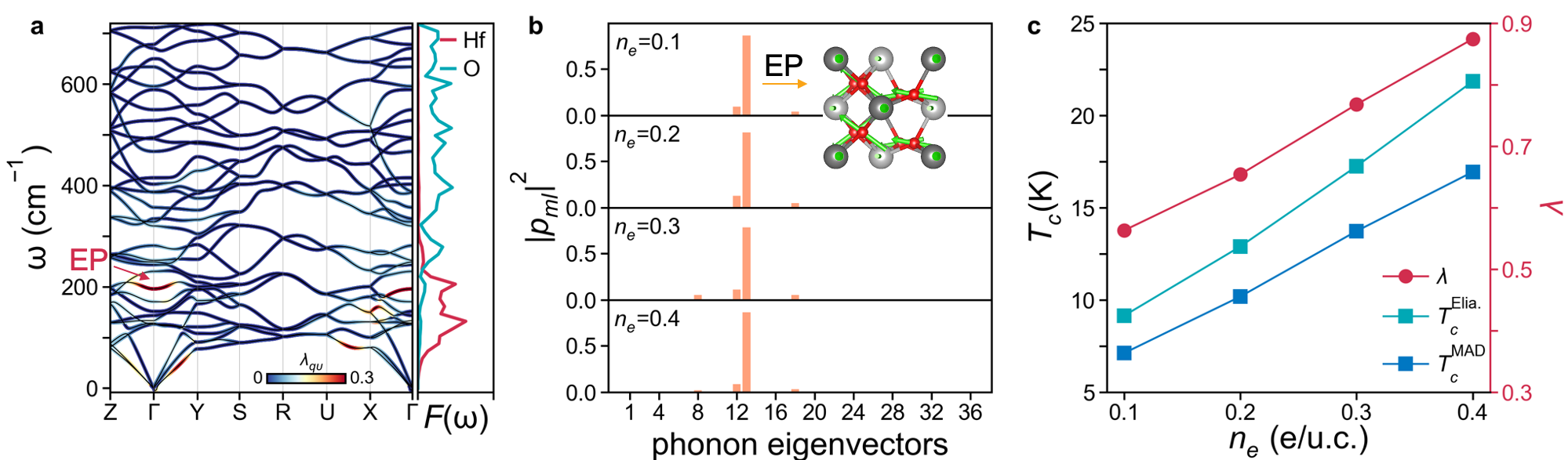}
 \caption{\textbf{a} $\lambda_{\mathbf{q}\nu}$-resolved phonon spectrum and projected phonon DOS of doped HfO$_2$ at $n_e=0.3$. The $\Gamma$-point nonpolar mode with the highest electron-phonon coupling strength is denoted as EP.
\textbf{b} Projection coefficients for EP at different doping concentrations. The inset shows the vibration modes of undoped EP.  \textbf{c} Superconducting transition temperature $T_c$ computed with two different methods ($T_c^{\rm MAD}$ and $T_c^{\rm Elia.}$) and total electron-phonon coupling constant $\lambda$ as a function of $n_e$. $\mu^*=0.1$ is used to estimate $T_c$.
 }
 \label{fig3}
 \end{figure}

We now focus on the electron-phonon coupling in electron-doped $Pca2_1$ HfO$_2$. Figure~\ref{fig3}\textbf{a} presents the mode-resolved electron-phonon coupling strength $\lambda_{\mathbf{q}\nu}$ for each individual phonon mode (indexed by $\mathbf{q}$ and band $\nu$) at $n_e$ = 0.3. Interestingly, the $\Gamma$-point phonon mode (denoted as EP) with the most significant electron-phonon coupling strength is a nonpolar optical mode. This feature again is distinct from that of electron-doped BaTiO$_3$, where the zone-center polar mode exhibits the greatest electron-phonon coupling strength~\cite{Ma21p2314}. The projected phonon DOS $F(\omega)$ (Fig.~\ref{fig3}\textbf{a}) reveals that the phonon modes with frequencies under $\approx$200 cm$^{-1}$ are predominantly governed by the vibrations of Hf atoms due to their larger atomic mass compared to O atoms.  Additionally, acoustic phonon bands along $\Gamma$-Y and U-X paths as well as nonpolar optical phonon bands along the $\Gamma$-X path have considerable magnitudes of $\lambda_{\mathbf{q}\nu}$.

The frequency of EP as a function of $n_e$ is displayed in Fig.~\ref{fig2}\textbf{b}, with the evolution determined using the mode-projection method, showing a notable mode softening with $\Delta \omega / \Delta n_e =-132$~cm$^{-1}$. Moreover, the phonon eigenvector of EP remains nearly invariant with respect to $n_e$. The projection coefficients of EP as a function of $n_e$ (Fig.~\ref{fig3}\textbf{b}) indicate that this mode mainly originates from the nonpolar optical mode of $m=13$ in undoped HfO$_2$, involving vibrations of both Hf and O atoms at roughly equivalent amplitudes (see inset of Fig.~\ref{fig3}\textbf{b}). This could explain the large electron-phonon coupling strength of EP: both the doped itinerant electrons and lattice vibrations are associated with Hf atoms.

The softening of EP induced by electron-doping further enhances its electron-phonon coupling strength as $\lambda_{\mathbf{q}\nu} \propto \omega_{\mathbf{q}\nu}^{-2}$.
The Eliashberg spectral function $\alpha^{2}F(\omega)$ and accumulative electron-phonon coupling $\lambda(\omega)$ of doped HfO$_2$ are provided in Supplemental Material.
The total electron-phonon coupling constant $\lambda$ at $n_e=0.3$ is 0.77, which is sufficiently large to induce phonon-mediated superconductivity~\cite{Ma21p2314,Sun21p045121,Zhao21p076301}.
The transition temperature $T_c$ is estimated with two methods, the McMillan-Allen-Dynes formula ($T^{\rm MAD}_c$)~\cite{Allen75p905} and the solution of the Migdal-Eliashberg equations ($T^{{\rm Elia.}}_c$)~\cite{Ponc16p116,Margine13p024505}.
As shown in Fig.~\ref{fig3}\textbf{c}, both $T^{\rm MAD}_c$ and $T^{{\rm Elia.}}_c$ elevate with increasing $n_e$.
It is well known that the magnitude of $T_c$ is sensitive to the value of Coulomb pseudopotential $\mu^{*}$~\cite{Richardson97p118,Lee95p1425}. Our test suggests that $T_c$ decreases from 31 K to 4 K as  $\mu^{*}$ varies from 0 to 0.3 at $n_e=0.3$ (see Supplemental Material).
While the precise value of superconducting $T_c$ is subject to uncertainty of $\mu^{*}$ and other parameter specifics, we believe that the enhancement of $T_c$ as $n_e$ increases, arising from the EP mode softening, is a robust outcome. A conservative predication for $n_e=0.3$ using $\mu^{*}=0.1$ is that ferroelectric HfO$_2$ transitions into a superconducting state when the temperature drops below $\approx$17~K.

\begin{figure}[ht]
\centering
\includegraphics[scale=0.5]{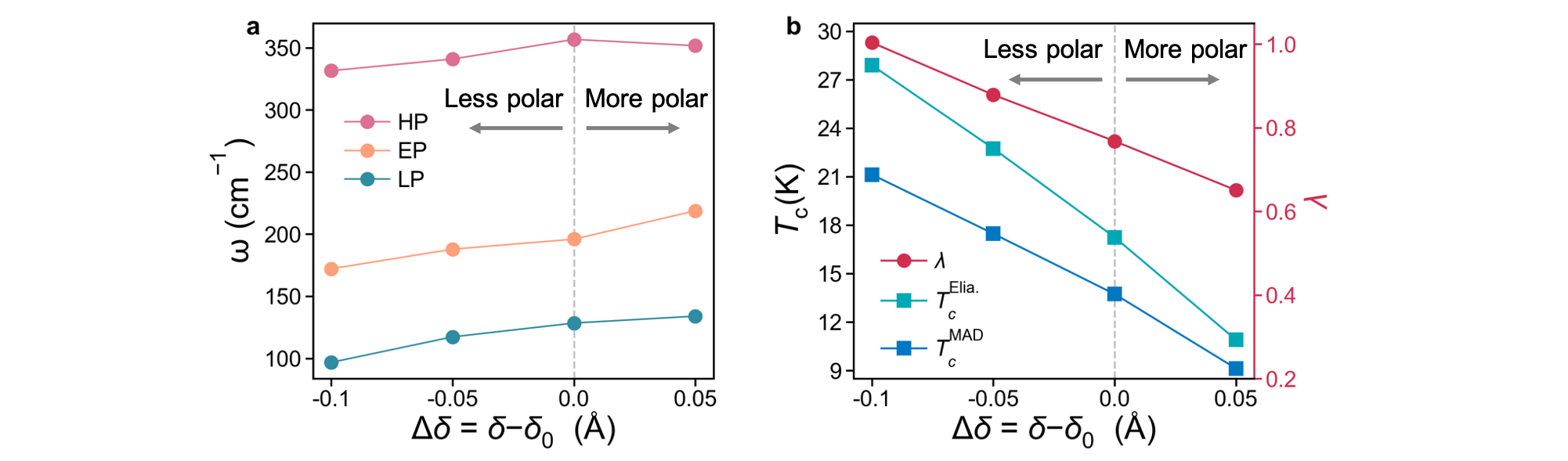}
\caption{\textbf{a} Frequencies of HP, EP, and LP modes at the $\Gamma$ point as a function of $\Delta \delta$ that measures the degree of inversion symmetry breaking ($P$). The doping concentration is $n_e=0.3$. \textbf{b} Superconducting transition temperature $T_c$ and total electron-phonon coupling constant $\lambda$ as a function $\Delta \delta$.}
  \label{fig4}
\end{figure}

Although the polar phonons (HP and LP) are not responsible for the emergence of superconductivity in electron-doped $Pca2_1$ HfO$_2$, we suggest a tunable $T_c$ could still be achieved by modulating the polar displacements of \OP~atoms, potentially enabling field-regulable quantum phases. The interplay between ferroelectricity and superconductivity arises from the fact that the displacement vector of \OP~atoms along the polar axis, activated by external stimuli such as epitaxial strain~\cite{Herrera19p124801,Ahadi19peaaw0120}, is not orthogonal to the eigenvector of EP. 
We perform a series of model calculations to quantify the relationship between $T_c$ and $\delta$ at a representative doping concentration ($n_e=0.3$). More specifically, we manually alter the positions of \OP~atoms; the change in the degree of inversion symmetry breaking is measured as $\Delta \delta = \delta - \delta_0 $, where $\delta_0$ represents the \OP~displacement at the ground state and a negative value of $\Delta \delta$ signifies reduced asymmetry. The relationship between $\Delta \delta$ and the frequency for LP, HP, and EP modes is depicted in  Fig.~\ref{fig4}\textbf{a}, revealing a positive correlation between $\omega$ and $\Delta \delta$. Consequently, pushing the structure to a less polar state will soften the EP mode, increase the electron-phonon coupling strength, and raise the $T_c$ (Fig.~\ref{fig4}\textbf{b}). By adjusting the positions of \OP~atoms within a range of $\pm0.1$~\AA~from their equilibrium values, it is possible to accomplish a shift in the superconducting temperature by a factor of two.

In summary, electron-doped $Pca2_1$ HfO$_2$ is identified as an Anderson-Blount ferroelectric metal where the polar phonon modes exhibit rather limited sensitivity to itinerant electrons. The emerging extreme charge-carrier-inert ferroelectricity provides a foundation for developing ferroelectric superconductors. A nonpolar optical phonon that undergoes softening due to electron-doping attains a sufficiently large electron-phonon coupling strength, thereby enabling BCS superconductivity. Notably, with an appropriate doping concentration, the superconducting $T_c$ of ferroelectric HfO$_2$ could surpass those of doped perovskite ferroelectrics ($\approx$2~K)~\cite{Ma21p2314} and most superconducting binary oxides such as LaO ($\approx$5~K)~\cite{Sun21p045121}, RuO$_2$ ($\approx$1.7~K) ~\cite{Ruf21p59,Masaki20p147001}, NbO$_x$ ($\approx$1.38~K), TiO$_x$ ($\approx$1~K)~\cite{Hulm72p291}, and SnO ($\approx$1.4~K)~\cite{Forthaus10p157001}. The distinct origins of ferroelectricity and superconductivity do not exclude their potential to interact, as the polar displacements of oxygen atoms involve both polar phonons that govern structural distortions and nonpolar phonons that mediate the binding of electrons. The prospect of a tunable $T_c$ over a window of at least 10 K, which is achievable by modulating polar distortions, presents exciting opportunities. We hope this work will stimulate experiments to explore superconductivity in electron-doped ferroelectric HfO$_2$ and to facilitate the design and discovery of silicon-compatible ferroelectric superconductors for novel device applications.  

\section{Acknowledgments}
XD and SL acknowledge the supports from National Key R\&D Program of China (2021YFA1202100), National Natural Science Foundation of China (12074319), and Westlake Education Foundation. The computational resource is provided by Westlake HPC Center.  

\bibliography{SL}
   \newpage

\newpage
 
\newpage

\newpage

\newpage

\newpage
\end{document}